# A new link between Boson and Fermion algebras: an alternative to supersymmetry

Philip Davies
Bournemouth and Poole College
pdavies@bpc.ac.uk

**Abstract**

*In this paper, we describe a new family of algebras which in the case of n = 2 reduces to the Fermion algebra and in the limiting case of n = ∞ reduces to the Boson algebra.*

**KEYWORDS**

*Fermion, Boson, supersymmetry. exclusion principle*

## 1. Introduction

The structure of the Fermion algebra, based originally on a postulate of Jordan and Wigner (1928), is that the Fermion annihilation and creation operators $b^\dagger$ and $b$ obey the anti-commutation relations

$[b, b^\dagger]_+ = 1$

This generates a finite algebra of dimension 2 and it is this relationship which ensures that states will not be multiply occupied providing the mathematical basis for the Pauli Exclusion Principle.

The structure of the Boson algebra is based on the equivalent commutation relation for the Boson annihilation and creation operators $a^\dagger$ and $a$

$[a, a^\dagger] = 1$

This generates an infinite dimensional algebra which ensures that Boson states can have multiple occupancy.

## 2. The Generalized Algebra

We define a generalized algebra with the following relations:

$[e, e^\dagger] = 1 - n/(n-1)!\ e^{\dagger n-1}\ e^{n-1}$

$[e^\dagger, e^\dagger] = [e, e] = 0$

$e^{\dagger n} = e^n = 0$

These relationships generate a finite algebra of dimension n which ensures that states will have a multiply occupancy of n-1 providing the mathematical basis for a generalized Pauli Exclusion Principle.

## 3. Special Cases

In the case of n = 2 the relationships result in the Fermion algebra.  In the case of

$$Lim_{n \to \infty} \frac{n}{(n-1)!} \to 0$$

the relationships result in the Boson algebra. The family of algebras therefore take the following form:

n = 1   $[e, e^\dagger] = 0$

n = 2   $[e, e^\dagger] = 1 - 2\, e^\dagger e$     (Fermion Algebra)

n = 3   $[e, e^\dagger] = 1 - 3/2\, e^{\dagger 2}\, e^2$

n = 4   $[e, e^\dagger] = 1 - 4/6\, e^{\dagger 3}\, e^3$

etc ............  ............................

n = ∞   $[e, e^\dagger] = 1$           (Boson Algebra)

## 4. Relationships common to the family of generalized algebras

In order to press home the similarities shared by these algebras we will summarize here those properties held in common by all the algebras including the Fermion and Boson algebras.

$e^\dagger e = N$   (the number operator)

$[e, e^\dagger e] = e$

$[e^\dagger, e^\dagger e] = -e^\dagger$

The generalized annihilation and creation operators have the following action upon the n dimensional ket space

$e \mid k > = \sqrt{k} \mid k-1 >$

$e^\dagger \mid k > = \sqrt{(k+1)} \mid k+1 >$

for k < n

By the repeated application of the n dimensional creation operator $e^\dagger$ upon the empty ket or vacuum state we find:

$e^{\dagger k} \mid 0 > = \sqrt{k!} \mid k >$

Since the creation operator is nilpotent of index n, the maximum value that the index k can take is

k = (n – 1) and thus a maximum of (n – 1) particles may be accommodated in the dynamical state. This is the generalized exclusion principle.

## 5. A generalized particle statistics

The class of distribution functions generated by the generalized algebra contains the special cases of the Fermion distribution (which applies to particles described by eigen functions which are antisymmetric with respect to the interchange of particle labels) and the Boson distribution (which applies to particles described by eigen functions which are symmetric with respect to the interchange of particle labels)

The average number of particles per state of energy E is given by

$$\overline{P} = \frac{\text{Total number of particles P}}{\text{Total number of states Z}}$$

In order to find the most probable configuration we use the variation principle and the method of Lagrange multipliers in the usual way, subject to a fixed number of particles and fixed total energy E

$$\overline{P} = \frac{\sum_{k=0}^{n-1} k e^{-k(Eb+c)}}{\sum_{k=0}^{n-1} e^{-k(Eb+c)}}$$

Where b and c are the Lagrange factors for the generalized algebra of dimension n.

For n = 2 this reduces to the Fermion case:

$$\overline{P} = \frac{1}{e^{(Eb+c)} + 1}$$

which we recognize as the standard Fermion distribution function.

For n → ∞ this reduces to the Boson case:

$$\overline{P} = \frac{1}{e^{(Eb+c)} - 1}$$

which we recognize as the standard Boson distribution function

## 6. Conclusion

We have demonstrated a new relationship between the Fermion and Boson algebras which sees them as two special cases of an infinite family of algebras. These generalized algebras describe particles which obey a generalized exclusion principle, limiting state occupation to (n-1) and obeying a generalized particle statistics intermediate between symmetric and antisymmetric.

Whether such particles exist in nature will need further investigation.

## 7. References

[1] Jordan and Wigner (1928) Uber das Paulische Aquivalenzverbot, in Selected Papers on Quantum Electrodynamics, Ed. J. Schwinger, Dover.

[2] Davies, P., (1981) The Weyl Algebra and an Algebraic Mechanics. Ph.D thesis, Birkbeck College, University of London.

[3] Weyl, H., (1931) The Theory of Groups and Quantum Mechanics, Reprinted by Dover Publications Inc., New York, USA (1950)